\documentclass[useAMS]{mn2e}
\usepackage{graphicx}

\begin{document}

\newcommand{\fe}{Fe~II}
\newcommand{\h}{H$_2$}
\newcommand{\ci}{C~I}

\title[On the excitation of the IR knots in the HH99 outflow]{On the
excitation of the infrared knots in the HH99 outflow
\thanks{Based on observations collected at the European Southern
Observatory, Paranal, Chile (70.C-138)}}
\author[C. M$^{\rm c}$Coey et al.]{C. M$^{\rm c}$Coey$^1$\thanks{E-mail:
carolyn.mccoey@durham.ac.uk},
T. Giannini$^2$,  D.R. Flower$^1$, A. Caratti o Garatti$^{2,3}$\\
$^1$ Physics Department, The University, Durham DH1 3LE, UK\\
$^2$ INAF-Osservatorio Astronomico di Roma, via Frascati 33, I-00040
Monteporzio Catone, Italy\\
$^3$ Universit\`a degli Studi 'Tor Vergata', via della Ricerca Scientifica
1, I-00133 Roma, Italy}

%
\date{Received date; Accepted date}
%

\maketitle

\begin{abstract} We present near infrared (IR) spectra (0.98--2.5 $\mu$m)  
from the group of Herbig-Haro (HH) objects comprising HH99: a series of
knots forming a bow, HH99B, and a separate knot, HH99A.  Observations of
{\h}, [{\fe}] and [{\ci}] are used to constrain shock model parameters and
determine the origin of the emission.  Previous work has shown that it is
likely that the atomic and ionic emission arises in regions of higher
ionization than the molecular emission. On the basis of observations, it
has often been suggested that the [{\fe}] and [{\ci}] emission could
arise, for example, at the apex of a bow shock, with the {\h} emission
produced in the wings.  Accordingly, we have combined models of
C-component and J-type shocks in order to reproduce the observed {\h},
[{\fe}] and [{\ci}] line spectra. We can account for the {\h} emission
towards the HH99B complex by means of J-type shocks with magnetic
precursors. We derive shock velocities of about 30 \mbox{km s$^{-1}$} and
ages of order 100 yr; the pre-shock gas has a density $n_\mathrm{H}$
$\approx$ 10$^4$ cm$^{-3}$ and the magnetic field $B \approx$ 100 $\mu$G.
The J-type shocks required to reproduce the intensities of the observed
[{\fe}] and [{\ci}] lines have velocities of 50 km s$^{-1}$. It is
necessary to assume that Fe has been previously eroded from grains,
probably by the earlier passage of a C-type shock wave. Thus, our analysis
supports the view that molecular outflows are episodic phenomena whose
observed emission arises in shock waves. \end{abstract}


\section{Introduction}

Bipolar outflows are produced during the accretion phase of low mass star
formation in molecular clouds.  The large scale outflows are observed in
the infrared domain and are traced, for example, by rotational transitions
of CO (e.g. $J = 3 \rightarrow 2$, $J = 2 \rightarrow 1$).  The outflows
often terminate in a bow-shaped structure and are marked by chains of
Herbig-Haro objects, which are condensations of warm gas observable in the
visible and in the near infrared. It is commonly thought that HH objects
are produced in episodic events and are: dense regions of the ambient
interstellar medium (ISM) which have been shocked by the passage of the
outflow; or ejecta from the forming star or its accretion disk; or the
consequence of instabilities within the outflow. The observed molecular
line shifts and widths and the presence of refractory species, such as Fe
and Si, with abundances higher than in the ambient medium, are indicative
of shock waves.  HH outflows have dynamical lifetimes of the order of
10$^3$--10$^5$ years (Bachiller, 1996), whereas the timescale for a shock
wave to attain steady state can be of the order of 10$^5$ years (Pineau
des For\^{e}ts, Flower \& Chi\`{e}ze, 1997; Chi\`{e}ze, Pineau des
For\^{e}ts \& Flower, 1998). It follows that non--equilibrium shock
modelling may be required when simulating the observed line intensities.

\begin{figure}
\centering
\includegraphics[width=9 cm]{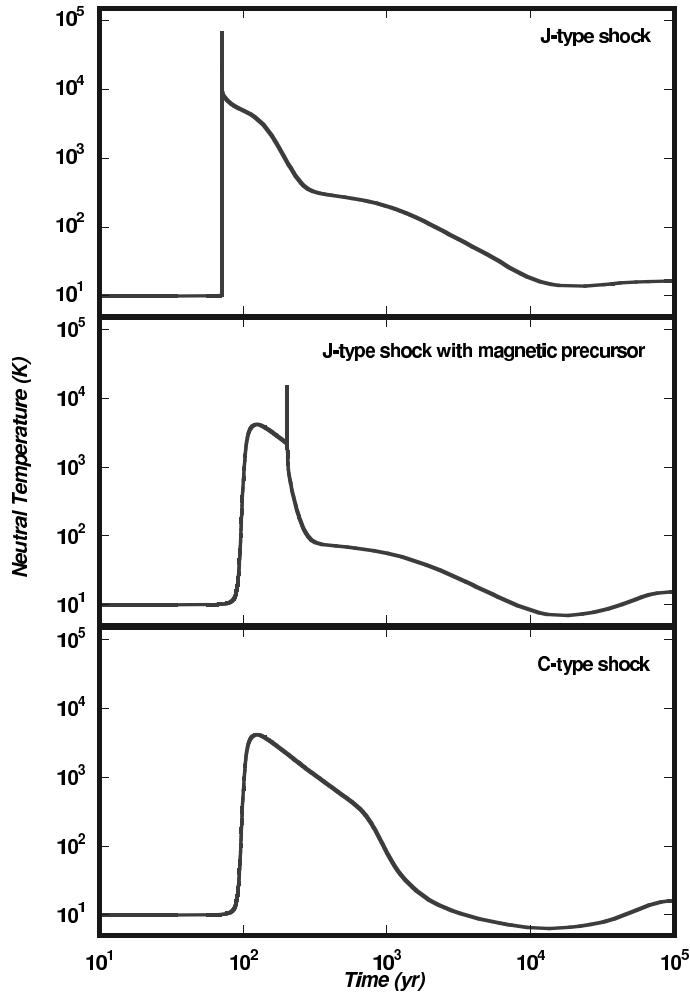}
\caption{Temperature profiles of J-type (top panel), J-type with magnetic
precursor (middle panel) and C-type (bottom panel) shock waves
($v_\mathrm{s}$=40 km s$^{-1}$, $n_\mathrm{H}$=10$^4$ cm$^{-3}$, $B$=100
$\mu$G). In evolutionary terms, a J-type shock with a magnetic precursor
can be considered to be intermediate between J-type and C-type shocks.
Note the increase in shock width and decrease in peak temperature in the
evolution from a J- to a C-type shock. }
\label{Profiles}
\end{figure}

Shock waves are usually classified as `jump' (`J-type') or continuous
(`C-type') (see Draine, 1980). Each type of shock wave gives rise to
different excitation conditions in the gas.  In a J-type shock of velocity
40 km s$^{-1}$, the temperature immediately behind the discontinuity
approaches 10$^5$~K, as compared with $T \approx 10$~K in the
preshock gas (see Figure~\ref{Profiles}). The gas density increases by no
more than a factor of 4 at the discontinuity, but the gas continues to be
compressed in the cooling flow. C-type shocks occur in regions of low
ionization, where the strength of the coupling between the charged and
neutral fluids is insufficient to prevent a velocity difference developing
between the two fluids.  The charged fluid and the magnetic field are
compressed ahead of the neutral fluid; this results in a broadening of the
shock wave and a decrease in the peak temperature (see Figure
\ref{Profiles}). In the presence of a sufficiently strong magnetic field,
an initially J-type shock may evolve into C-type. At early times, a J-type
discontinuity remains embedded in the C-type flow (Pineau des For\^{e}ts,
Flower \& Chi\`{e}ze, 1997; Smith \& Mac Low, 1997; Chi\`{e}ze, Pineau des
For\^{e}ts \& Flower, 1998; Lesaffre, Chi\`{e}ze, Cabrit \& Pineau des
For\^{e}ts, 2004a,b). These `intermediate' types of
shock wave are referred to as `J-type with a magnetic precursor'; they are
narrower than
the equivalent steady--state C-type shock waves and have higher maximum
temperatures.

The large number of IR rovibrational emission lines of {\h} observed
towards HH objects suggests the importance of {\h} as a coolant.  The
excitation diagrams derived from these lines can be used to constrain the
model parameters, such as the shock velocity and age, and the pre-shock
gas density.  Comparison of the observed {\h} emission with the
predictions of shock models has shown the best agreement for shock waves
which are C-type or have a C-component, i.e. J-type with a magnetic
precursor (Giannini et al., 2004; Flower et al., 2003; Le Bourlot et al.,
2002).

In addition to being observable in rovibrational transitions of {\h}, HH
outflows can be traced through their emission in high-$J$ CO and other
molecular lines and the emission of atoms and ions (e.g. Liseau,
Ceccarelli \& Larsson, 1996; Benedettini et al., 2000). With increasing
shock speed, the temperature and density of the gas in the shock wave
increase; {\h} is dissociated and H begins to be ionized. As a
consequence, the contribution of {\h} to the radiative cooling decreases
and that of other species, such as H$_{2}$O, CO, O, C and Fe$^{+}$,
assumes greater importance. The strongest ionic and atomic lines observed
are those of [{\fe}] and [{\ci}]. In a C-type shock wave, or in a magnetic
precursor to a J-type shock, charged grains can be eroded by neutral
particles, owing to the relative streaming of the charged and neutral
fluids. At sufficiently high shock speeds, the (refractory) grain cores
can be partially eroded, releasing elements such as Fe and C into the gas
phase; see May et al. (2000).  Neutral iron is ionized rapidly in the gas
phase, predominantly through charge transfer reactions with ions, such as
H$_{3}$$^{+}$ and H$_{3}$O$^{+}$, which have larger ionization potentials.
Fe$^{+}$ may then be excited collisionally and emit the [{\fe}] forbidden
line spectrum.

Giannini et al. (2004) showed that it is not possible to reproduce the
atomic and ionic emission observed in HH objects with the same planar model
as is
deduced from the {\h} emission; the atomic and ionic emission requires a
higher degree of ionization.  Several possible explanations have been
proposed for the presence of ionized
species, such as [{\fe}], in conjunction with {\h}. For example,  Gredel
(1994, 1996) suggested curved J-type shocks, while other authors have
proposed a combination of C- and J-type shocks (Molinari et al., 2000) or
J-type shocks with magnetic precursors (Hartigan, Curiel \& Raymond, 1989).
Some observed features of HH objects, such as the line widths and range of
excitation conditions inferred from {\h} observations (e.g. Eisl\"{o}ffel,
Smith \& Davis, 2000) and the existence of `caps' of [{\fe}] emission ahead
of that of {\h} (e.g. Davis, Smith \& Eisl\"{o}ffel, 2000), have led to
suggestions that the {\h} emission arises
in the wings the and atomic and ionic emission originates at the apex of a bow
shock.  Alternatively, the atomic and ionic emission might arise in a
reverse shock, in the jet, and the {\h} emission in the molecular gas
which has been impacted by the jet.  In either case, the degree of
ionization must be sufficient for electron collisions to dominate the
excitation of the atomic and ionic emission lines; such conditions can be
realized in J-type shocks.

We identified as a good observational case to test shock models the
HH99 region ($\alpha$ = 19$^{h}$02$^{m}$05.4$^{s}$, $\delta$ =
-36$\degr$54$\arcmin$40$\arcsec$ (J2000.0)), a Herbig-Haro complex in the
RCrA molecular core ($d$ = 130 pc: Marraco \& Rydgren, 1981). This complex
comprises a prototype of bow shock (namely HH99B) and a bright wing
feature (knot A), probably originating in the interaction of the flow with
a dense ambient clump or cloudlet (Davis et al., 1999). On the basis of
optical images and high resolution spectra, Hartigan \& Graham (1987) have
suggested that the HH99 complex is the redshifted lobe of the HH100 flow,
powered by the embedded protostar HH100 IR. More recently, near infrared
images have demonstrated that the bow shock/jet define a direction
consistent with the position of the infrared source IRS 9 and the pre-main
sequence star RCrA (Wilking et al., 1997). The low resolution K-spectrum
of HH99B3 and echelle spectra of the H$_2$ 2.122 $\mu$m line have been
obtained by Davis et al. (1999).

In order to better probe the temperature along the shocked gas and to
constrain observationally the shock model parameters, we have investigated
spectroscopically the HH99 complex by observing the complete (0.98 -- 2.5
$\mu$m) near infrared spectrum across the bow and along the wing (see Figure
\ref{HH99:fig}). We present the observations of HH99 in Section \ref{Obs}
and discuss the model predictions in Section \ref{Model}, where we also
give a brief description of the shock code which was used.  We summarize
our findings in Section \ref{Conc}.

\section{Observations and data reduction}\label{Obs}
\begin{figure}
\centering
\includegraphics [width=9 cm] {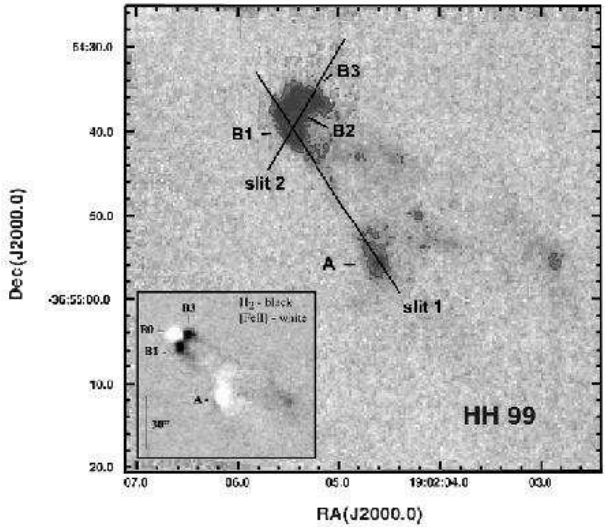}
\caption{HH99: H$_2$ (2.122 $\mu$m) image (continuum not subtracted)
obtained at ESO-NTT with Sofi IR camera. Knot labels and slit positions
are superposed.  The inset shows Figure~1b of Davis et al. (1999), on which
we have also marked the knot B0.
\label{HH99:fig}}
\end{figure}

The spectroscopic observations of HH99 were made in July 2002, at the
ESO-VLT, with the ISAAC near infrared camera and at a resolution of
0.147$\arcsec$/pixel (Cuby et al., 2003). In order to investigate the head
and wings of the bow shock, long--slit spectra were obtained for two
different position angles (P.A. = 32.4$\degr$ and 329.5$\degr$, referred
to as slit 1 and slit 2 respectively; see Figure~\ref{HH99:fig}) in four
segments, centered on the bands $z$, $J$, $H$ and $K$, covering a spectral
range from 0.98 to 2.50 $\mu$m. With reference to the nomenclature adopted
by Davis et al. (1999), slit 1 encompasses knots A, B1 and a further zone
of [{\fe}] emission, visible in their Figure~1, to which we shall
refer as B0. Slit 2 encompasses knots B1, B2 and B3. The slit width was
0.6$\arcsec$, corresponding to a spectral resolution from about 750 to
900, moving from the shorter to the longer wavelengths. To perform our
spectroscopic measurements, we adopted the usual ABB'A' configuration, for
a total integration time of 400~s per band. All the raw data were reduced
by using the IRAF\footnote{IRAF (Image Reduction and Analysis Facility) is
distributed by the National Optical Astronomy Observatories, which are
operated by AURA, Inc., under a cooperative agreement with the National
Science Foundation.} package, applying standard procedures for sky
subtraction, flat--fielding and bad pixel removal. Observations were
sky subtracted and corrected for the curvature, due to optical
distortions, by fitting sky lines row-by-row. Atmospheric features
were removed by dividing the
spectra by a telluric standard star (O spectral type), normalized to the
blackbody function at the star's temperature and corrected for hydrogen
recombination absorption features intrinsic to the star. The wavelength
calibration was obtained using sky OH lines (Rousselot et al., 2000).

\begin{figure*}
 \centering
   \includegraphics [width=12 cm] {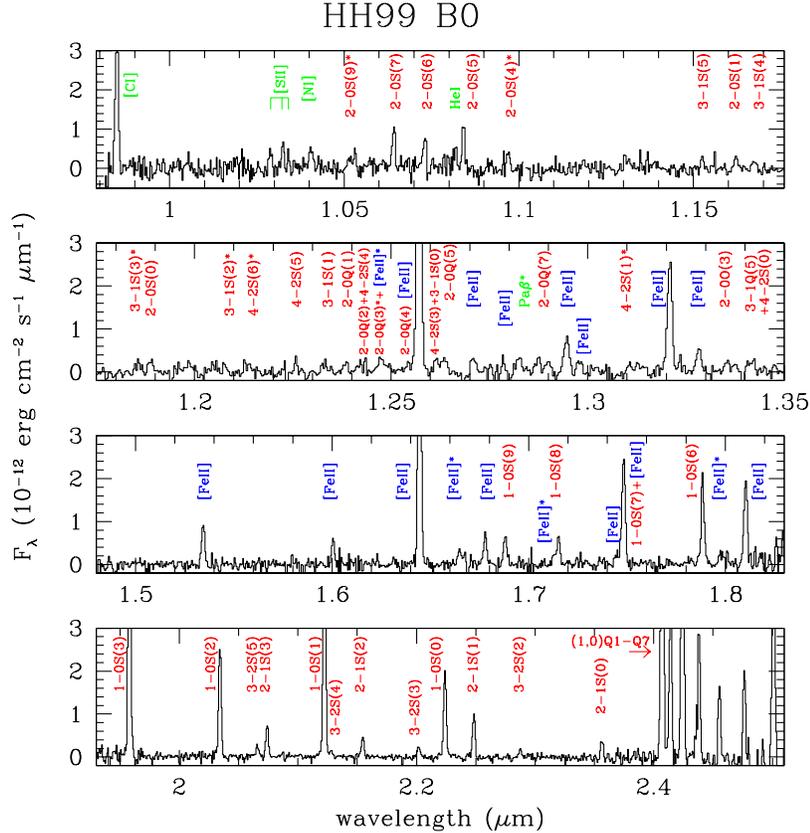}
   \caption{ 0.95--2.5 $\mu$m low resolution spectrum of knot B0 in HH99.
   An asterisk near the line identification marks detections between 2 and
3$\sigma$.
\label{HH99_B0_sp.eps:fig}}
\end{figure*}

All the observed spectra are rich both in H$_2$ (vibrational quantum
number $1 \le v \le 4$) and ionic lines, as observed in other HH objects
(Nisini et al., 2002). Figure~\ref{HH99_B0_sp.eps:fig} shows the spectrum
of the knot B0, where the most prominent ionic features are detected,
mainly [\fe] lines emitted by the first 13 energy levels. The other ionic
features detected in this knot are the [\ci] doublet at 0.983 and 0.985
$\mu$m and, at a lower S/N, the [S~II] 1.029, 1.032 and 1.034 $\mu$m
transitions, two [N~I] blended doublets (1.040/1.041 $\mu$m) and the He~I
1.083 $\mu$m recombination line. Pa$\beta$ is also just visible at \mbox{S/N
$\leq$ 3}.

The spectra of the other knots observed present H$_2$ (for \mbox{$v \le  3$}),
[\ci] and [\fe] transitions, but other ionic features are weak or absent,
probably indicating lower excitation conditions along the bow wings.

Line fluxes have been obtained by fitting the line profile with a single
or, in the case of blending, a double gaussian. Flux calibration
(associated uncertainty: 10$\%$) was achieved by observing the photometric
standard star HR7136 in the $J$, $H$ and $K$ bands; the photometric data
were taken from the catalogue of Van der Bliek et al. (1996). Ratios of
lines falling in different bands have an uncertainty of about 15$\%$, due
mainly to seeing variations during the night; this error was estimated by
measuring the intensities of lines present in two adjacent bands.

The complete list of the line fluxes is available only in the electronic
version of this paper. In Tables~\ref{fluxmola} and \ref{fluxmolb}, we
report, as examples, the lines identified in the B0 spectrum (column 1),
together with the vacuum wavelengths (column 2) and the measured fluxes
(column 3). The uncertainties associated with these data derive only from
the rms of the local baseline, multiplied by the linewidths (which are
always comparable with the width of the instrumental profile). Our line
fluxes are systematically lower than indicated by the photometry of Davis
et al. (1999). We believe that these discrepancies are attributable to the
fact that the areas of emission covered by our slits are from three to ten
times smaller than the dimensions of the knots. However, the lines fluxes
measured in knot B3 are in good agreement with those derived from low
resolution spectroscopy by Davis et al. (1999), if the different slit
widths and orientations are taken into account.

The use of ratios of intensities of pairs of lines emitted by a common
upper level has long been recognized as an important tool for deriving
extinction in HH objects (see, for example, Gredel 1994). In the spectra
of HH99 knots, lines fulfilling this criterion come both from [\fe] and
H$_2$, and this circumstance enables to make two independent
determinations of the visual extinction, $A_\mathrm{V}$. From the ratio of
the intensities of the 1.257$\mu$m and 1.644$\mu$m lines, which are the
[\fe] lines observed with the highest signal-to-noise ratio, we derive $4
\le A_\mathrm{V} \le 10$ mag (see Table \ref{fluxtab}), when the reddening
law of Rieke \& Lebofsky (1985) is adopted. Values of $A_\mathrm{V}$ which
are systematically lower, by about 2 to 3 mag, are obtained when using
pairs of H$_2$ transitions [such as 2-0 S(i)/2-0 Q(i+2) and 3-1 S(i)/3-2
S(i)]. This result, which has been found in other Herbig-Haro
condensations (e.g. Giannini et al. 2004, Nisini et al. 2002), shows that
the iron lines arise in regions which are more deeply embedded than those
responsible for the molecular hydrogen emission.

%
\begin{table*}
\caption[]{H$_2$ lines observed in HH99B0.}
\label{fluxmola}
\begin{tabular}{ccc}
\hline\\[-5pt]
Term                                                              &
$\lambda$ ($\mu$m)   & $F\pm\Delta~F$  (10$^{-16}$erg\,cm$^{-2}$\,s$^{-1}$)
\\
\hline\\[-5pt]
 ~2--0 S(9)                                                       & 1.053
&   4.9 $\pm$ 2.0$^a$ \\
 ~2--0 S(7)	                                                  & 1.064
	          &  13.4 $\pm$ 2.7     \\
 ~2--0 S(6)	                                                  & 1.073
	          &   7.4 $\pm$ 1.9     \\
 ~2--0 S(5)	                                                  & 1.085
	          &  17.6 $\pm$ 2.3     \\
 ~2--0 S(4)                                                       & 1.100
&   9.1 $\pm$ 4.0$^a$ \\
 ~3--1 S(5)	                                                  & 1.152
	          &   6.4 $\pm$ 1.3     \\
 ~2--0 S(1)	                                                  & 1.162
	          &   5.6 $\pm$ 1.2     \\
 ~3--1 S(4)	                                                  & 1.167
	          &   3.7 $\pm$ 1.2     \\
 ~3--1 S(3)                                                       & 1.186
&   5.4 $\pm$ 2.0$^a$ \\
 ~2--0 S(0)	                                                  & 1.189
	          &   3.6 $\pm$ 1.1     \\
 ~3--1 S(2)                                                       & 1.207
&   3.4 $\pm$ 1.4$^a$ \\
 ~4--2 S(6)                                                       & 1.214
&   3.2 $\pm$ 1.2$^a$ \\
 ~4--2 S(5)	                                                  & 1.226
	          &   3.9 $\pm$ 1.2     \\
 ~3--1 S(1)	                                                  & 1.233
	          &   5.7 $\pm$ 0.7     \\
 ~2--0 Q(1)	                                                  & 1.238
	          &   4.6 $\pm$ 1.0     \\
 ~2--0 Q(2)+4-2S(4)                                               &
1.242+1.242           &   3.6 $\pm$ 0.8     \\
 ~2--0 Q(3)+~[\fe]\,$a^4\!D_{3/2}-a^6\!D_{5/2}$                   &
1.247+1.248           &   4.0 $\pm$ 1.4$^a$ \\
 ~2--0 Q(4)	                                                  & 1.254
	          &   2.9 $\pm$ 0.6     \\
 ~4--2 S(3)+3-1S(0)                                               &
1.261+1.262           &   6.7 $\pm$ 1.2     \\
 ~2--0 Q(5)	                                                  & 1.263
	          &   5.2 $\pm$ 1.7     \\
 ~2--0 Q(7)	                                                  & 1.287
	          &   5.2 $\pm$ 0.9     \\
 ~4--2 S(1)                                                       & 1.311
&   5.0 $\pm$ 1.8$^a$ \\
 ~2--0 O(3)	                                                  & 1.335
	          &   5.7 $\pm$ 1.3     \\
 ~3--1 Q(5)+4-2S(0)                                               &
1.342+1.342           &   5.9 $\pm$ 1.2     \\
 ~1--0 S(9)	                                                  & 1.688
	          &  14.2 $\pm$ 2.0     \\
 ~1--0 S(8)	                                                  & 1.715
	          &  12.1 $\pm$ 1.8     \\
 ~1--0 S(7)+~[\fe]\,$a^4\!P_{3/2}-a^4\!D_{7/2}$	                  &
1.748+1.749           &  47.5 $\pm$ 1.4     \\
 ~1--0 S(6)	                                                  & 1.788
	          &  34.6 $\pm$ 1.3     \\
 ~1--0 S(3)	                                                  & 1.958
	          & 199.6 $\pm$ 6.5     \\
 ~1--0 S(2)	                                                  & 2.034
	          &  61.0 $\pm$ 1.8     \\
 ~3--2 S(5)	                                                  & 2.066
	          &   7.2 $\pm$ 1.4     \\
 ~2--1 S(3)	                                                  & 2.073
	          &  18.3 $\pm$ 1.4     \\
 ~1--0 S(1)	                                                  & 2.122
	          & 147.4 $\pm$ 1.4     \\
 ~3--2 S(4)	                                                  & 2.127
	          &   4.5 $\pm$ 0.7     \\
 ~2--1 S(2)	                                                  & 2.154
	          &  13.1 $\pm$ 1.4     \\
 ~3--2 S(3)	                                                  & 2.201
	          &   8.5 $\pm$ 1.4     \\
 ~1--0 S(0)	                                                  & 2.223
	          &  48.8 $\pm$ 1.1     \\
 ~2--1 S(1)	                                                  & 2.248
	          &  26.6 $\pm$ 1.2     \\
 ~3--2 S(2)	                                                  & 2.287
	          &   7.0 $\pm$ 1.2     \\
 ~2--1 S(0)	                                                  & 2.355
	          &  14.2 $\pm$ 1.3     \\
 ~1--0 Q(1)	                                                  & 2.407
	          & 212.4 $\pm$ 4.3     \\
 ~1--0 Q(2)	                                                  & 2.413
	          &  84.7 $\pm$ 4.3     \\
 ~1--0 Q(3)	                                                  & 2.424
	          & 209.5 $\pm$ 4.3     \\
 ~1--0 Q(4)	                                                  & 2.437
	          &  67.1 $\pm$ 4.1     \\
 ~1--0 Q(5)	                                                  & 2.455
	          &  54.5 $\pm$ 4.9     \\
 ~1--0 Q(6)	                                                  & 2.476
	          &  48.9 $\pm$11.7     \\
 ~1--0 Q(7)	                                                  & 2.500
	          & 140.8 $\pm$15.0     \\
\hline\\[-5pt]
Notes: $^{a}$S/N between 2 and 3.\\
\end{tabular}
\end{table*}

%
\begin{table*}
\caption[]{Atomic and Ionic lines observed in HH99B0.}
\label{fluxmolb}
\begin{tabular}{ccc}
\hline\\[-5pt]
Term
&  $\lambda$ ($\mu$m)     & $F\pm\Delta~F$
(10$^{-16}$erg\,cm$^{-2}$\,s$^{-1}$) \\
\hline\\[-5pt]
~[\ci]\,$^1\!D_{2}-^3\!P_{1}$+~[\ci]\,$^1\!D_{2}-^3\!P_{2}$
& 0.983+0.985             &  66.1  $\pm$ 5.6     \\
~[S~II]\,$^2\!P_{3/2}-^2\!D_{3/2}$
& 1.029	                &  10.3  $\pm$ 1.7     \\
~[S~II]\,$^2\!P_{3/2}-^2\!D_{5/2}$
& 1.032	                &   9.5  $\pm$ 1.2     \\
~[S~II]\,$^2\!P_{3/2}-^2\!D_{3/2}$
& 1.034	                &   5.6  $\pm$ 1.0     \\
~[N~I]\,$^2\!P_{3/2,1/2}-^2\!D_{5/2}$ +$^2\!P_{3/2,1/2}-^2\!D_{3/2}$
& 1.040-1.041             &  11.4  $\pm$ 2.2     \\
~He~I\,$^3\!S_{1}-^3P^o_{0,1,2}$
& 1.083	                &   7.9  $\pm$ 2.1     \\
~[\fe]\,$a^4\!D_{7/2}-a^6\!D_{9/2}$
& 1.257	                & 119.1  $\pm$ 1.0     \\
~[\fe]\,$a^4\!D_{1/2}-a^6\!D_{1/2}$
& 1.271	                &   8.1  $\pm$ 1.7     \\
~[\fe]\,$a^4\!D_{3/2}-a^6\!D_{3/2}$
& 1.279	                &  10.5  $\pm$ 2.9     \\
~Pa$\beta$
	      & 1.282                   &   4.8  $\pm$ 2.0$^a$ \\
~[\fe]\,$a^4\!D_{5/2}-a^6\!D_{5/2}$
& 1.295	                &  16.6  $\pm$ 1.8     \\
~[\fe]\,$a^4\!D_{3/2}-a^6\!D_{1/2}$
& 1.298	                &   4.7  $\pm$ 1.0     \\
~[\fe]\,$a^4\!D_{7/2}-a^6\!D_{7/2}$
& 1.321	                &  41.0  $\pm$ 1.8     \\
~[\fe]\,$a^4\!D_{5/2}-a^6\!D_{3/2}$
& 1.328	                &  10.3  $\pm$ 2.0     \\
~[\fe]\,$a^4\!D_{5/2}-a^4\!F_{9/2}$
& 1.534	                &  19.4  $\pm$ 1.5     \\
~[\fe]\,$a^4\!D_{3/2}-a^4\!F_{7/2}$
& 1.600	                &  11.0  $\pm$ 2.1     \\
~[\fe]\,$a^4\!D_{7/2}-a^4\!F_{9/2}$
& 1.644	                & 133.9  $\pm$ 1.5     \\
~[\fe]\,$a^4\!D_{5/2}-a^4\!F_{7/2}$
& 1.678	                &  12.3  $\pm$ 2.4     \\
~[\fe]\,$a^4\!D_{3/2}-a^4\!F_{5/2}$
& 1.711                   &   5.2  $\pm$ 1.8$^a$ \\
~[\fe]\,$a^4\!D_{1/2}-a^4\!F_{3/2}$
& 1.745	                &   8.2  $\pm$ 1.9     \\
~[\fe]\,$a^4\!D_{3/2}-a^4\!F_{3/2}$
& 1.798                   &   6.4  $\pm$ 2.7$^a$ \\
~[\fe]\,$a^4\!D_{7/2}-a^4\!F_{7/2}$+~[\fe]\,$a^4\!P_{5/2}-a^4\!D_{7/2}$
& 1.810+1.811             &  36.9  $\pm$ 3.8     \\
\hline\\[-5pt]
Notes: $^{a}$S/N between 2 and 3.\\

\end{tabular}

\end{table*}

\section{Comparison with predictions of shock models}\label{Model}
In this Section, we compare the predictions of the shock models with the {\h},
[{\fe}] and [{\ci}] emission observed towards HH99. Relevant aspects of
the modelling of the emission are summarized in Section \ref{modcon}, and
the results are presented and discussed in Section \ref{modres}.

\subsection{The model}\label{modcon}

\subsubsection{Theory}

The features of the shock model are described by Le Bourlot et al. (2002)
and Flower et al. (2003). Here, we review only those aspects relevant to
the present study. The code, \textsc{MHD$\_$VODE}, solves for
one--dimensional, planar, multi--fluid flow and can simulate not only
steady state J- and C- type shock waves, but also non--equilibrium J-type
shocks with magnetic precursors. The differential equations which
determine the abundances of the chemical species are solved in parallel
with the magnetohydrodynamical conservation equations for the neutral,
positively and negatively charged fluids. The chemical network consisted
of over 1000 reactions involving 138 species.

The method used to calculate the emission by molecular hydrogen was
explained by Giannini et al. (2004). The processes which populate and
depopulate the rovibrational levels of {\h} are: collisional excitation
and de-excitation; spontaneous radiative decay; collisional dissociation
and ionization; and reformation of {\h} on grains, which occurs in the
wake of the shock wave. We calculate the distribution of population among
150 rovibrational levels of {\h} (i.e. up to an energy of 3.9$\times 10^4$
K).  The equations for the level populations are solved in parallel with
the chemical and dynamical conservation equations. This approach is
essential to ensure the accuracy of the computed {\h} column densities,
because the level populations do not respond instantaneously to changes in
the physical state of the gas. Radiative pumping of {\h} is considered to
be unimportant compared with collisional excitation within a shock wave
(see Giannini et al., 2004) and is not taken into account.

The {\h} emission line intensities predicted by the shock model can be
used to generate an excitation diagram (also known as a Boltzmann plot);
this is a plot of ${\rm ln}(N_{v J}/g_J)$ against $E_{v J}/k$, where $N_{v
J}$ (cm$^{-2}$) is the column density of level ($v, J$), $E_{v J}/k$ (K)
is its excitation energy, and $g_J = (2J+1)(2I+1)$ is its statistical
weight. The nuclear spin quantum number $I = 1$ for ortho--H$_2$ and $I =
0$ for para--H$_2$. Under conditions of thermodynamic equilibrium, this
plot is a straight line whose gradient is inversely proportional to the
kinetic temperature of the gas.  In a shock wave, on the other hand, there
is a range of kinetic temperatures. If the gas is in local thermodynamic
equilibrium (LTE), the Boltzmann plot is a smooth curve, with little
scatter of the points about the median line.  Departures from LTE enhance
the scatter. Thus, the excitation diagram is a useful diagnostic tool when
studying H$_2$ emission lines from shocked molecular gas.

The rates of the collisions which excite vibrationally the {\h} molecules
(predominantly those with H) increase with the gas density, $n_{\rm H}$,
and the shock velocity, $v_s$. Hence, departures from LTE tend to decrease
with increasing values of $n_{\rm H}$ and $v_s$.  The evolutionary age of
the shock wave affects differentially the populations of the more highly
excited levels of {\h}. The younger the object, the closer is the shock
wave to being pure J-type and the higher are the column densities of the
more highly excited rovibrational levels.  Furthermore, the populations of
these levels tend to be closer to LTE.  A higher degree of ionization of
the gas (owing to the presence of a background UV radiation field, for
example) leads to an increase in the strength of the coupling between the
neutral and the charged fluids; this results in a narrower shock wave,
nearer to J- than to C-type, and so in this case also the {\h} populations
tend to be closer to LTE.  In a J-type shock with a magnetic precursor,
the levels of $v = 0, 1$ are populated principally in the precursor and
exhibit greater departures from LTE than the higher levels. Owing to
radiative cascade from excited vibrational states, most of the population
remains in the $v = 0$ ground vibrational state. Furthermore, within each
vibrational manifold, population tends to accumulate in the lower
rotational levels, following rotationally inelastic collisions, mainly
with H$_2$ and He (Flower et al., 2003).

\begin{figure}
\centering
\includegraphics [width=8.5 cm] {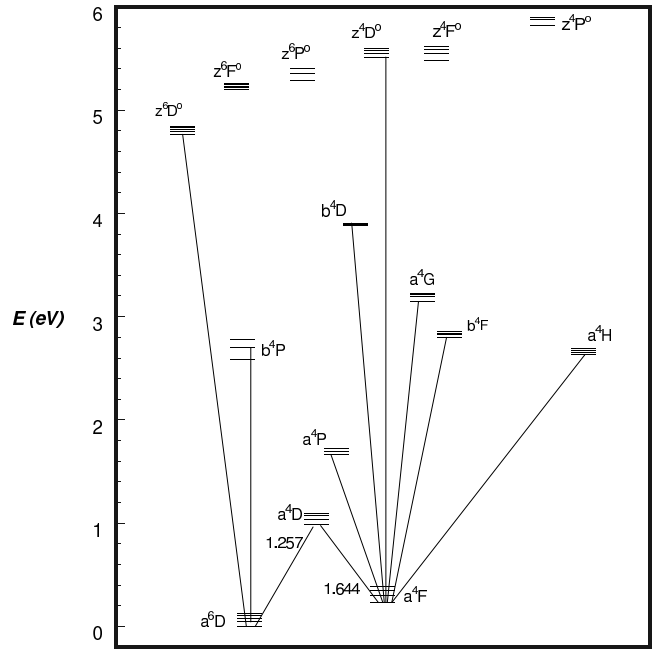}
\caption{Energy level diagram for Fe$^+$.  The strongest transition from
each of the even parity terms is marked.  The odd parity terms decay most
strongly to terms with the same multiplicity and the strongest transitions
from the z$^6$D$^{\rm o}$ and z$^4$D$^{\rm o}$ terms are marked as
examples.  The
strongest observed IR transitions, the 1.257 and 1.644 $\mu$m lines,
are also indicated.
\label{Felev}}
\end{figure}

If {\h} is dissociated to H in a J-type shock wave, cooling by atoms and
ions, such as C, O, and Fe$^+$, can become relatively more important. The
emission from (and contribution to the cooling by) these and other
important species were computed by the model. In the case of atomic C, we
included collisional and radiative transitions between the five energy
levels arising from the $^3$P, $^1$D and $^1$S terms using data from the
compilation of Mendoza (1983). The model of Fe$^{+}$ included transitions
among the 19 energy levels that arise from the a$^6$D, a$^4$F, a$^4$D,
a$^4$P and b$^{4}$P terms. The spontaneous radiative transition
probabilities computed by Nussbaumer \& Storey (1988) were used, but, as
their data extend only to the a$^4$P term, the results of Quinet, Le
Dourneuf \& Zeippen (1996) were adopted for the b$^4$P term. Giannini et
al. (2004) have established that collisions with electrons are likely to
dominate the excitation of Fe$^+$.  The corresponding electron collision
strengths were taken from Zhang \& Pradhan (1995).

All of the observed [{\fe}] IR transitions are from the a$^4$D term, which
lies at an energy of approximately 1 eV (10$^4$ K) above the ground state:
see Figure~\ref{Felev}.  In order to calculate correctly the populations
of the levels of this and other terms, it is necessary to ensure that all
significant cascade effects have been included.  The highest level of the
five terms considered lies at an energy of approximately 3 eV (3$\times
10^4$ K), which is lower than the peak temperatures (of order 10$^5$~K)
attained in the J-type shock models.  The inclusion of higher energy
terms, and in particular the a$^4$H, b$^4$F, a$^4$G and b$^4$D terms, which
decay preferentially to the a$^4$F term, could conceivably change the
distribution of population in the a$^4$D term.  As a check on the
convergence of our results, the 50 km s$^{-1}$ J-type shock model
considered in Section \ref{modres} below was rerun, including all of the
142 energy levels for which Zhang \& Pradhan (1995) provided data, i.e.
levels up to 11.5 eV \mbox{(1.3 $\times$ 10$^5$ K)} above ground. The set
of radiative transition probabilities was extended by means of the data of
Quinet et al. (1996), up to the b$^4$D term, and of Nahar (1995) for the
odd parity terms. The overall effect of including these terms is small:
the intensities of the [Fe II] lines which are observed were
reduced, but by no more than 10$\%$, which is to be compared with the
estimated error of 20$\%$ in the observed line intensities.  We conclude
that the basis of 19 levels is sufficient to model the [{\fe}] emission
lines which are observed, given the current observational uncertainties.

\subsubsection{Application}

As noted by Davis et al. (1999), and as can be seen in Figure
\ref{HH99:fig}, the morphology of the HH99B complex suggests a bow shock.
A bow shock can be modelled approximately as a sequence of planar shock
fronts, for which the effective shock speed varies with the angle of
attack.  In the present study, we assume a dissociative J-type shock to
occur at the apex of the bow, and J-type shocks with magnetic precursors
to dominate in the wings. Each shock model is computed independently of
the others, but we consider, in Section \ref{h2feemiss}, the compatibility
of the observations with the sum of the H$_2$ emission from the different
components. We found that a two--component model, comprising a pure J-type
shock and a J-type shock with a magnetic precursor, of lower speed, is
able to account for the observations of both H$_2$ and of the atomic and
ionic emission lines.  More sophisticated treatments of the geometry of
bow shocks have been undertaken (e.g. Smith, Brand \& Moorhouse, 1990;
Smith, Khanzadyan \& Davis, 2003).  However, it has proved too
computationally expensive to include in such models a proper treatment of
the chemical and molecular processes that influence shock waves and which
are considered here.

The optimal values of the shock speed and pre-shock gas density (in the
range $10^{3} \le n_{\mathrm{H}} \le 10^{6}$ cm$^{-3}$) were determined by
comparing calculated with observed {\h} excitation diagrams and [Fe II]
line intensities. The scaling relation B($\mu$G)$ = b
[n_\mathrm{H}$(cm$^{-3}$)]$^{0.5}$ determined the initial value of the
magnetic induction. Ideally, $b$ should be determined observationally, but
this is not generally feasible.  In this paper we use the value of $b$=1,
consistent with equipartition of energy.

The effect of photoionization processes on the degree of ionization in the
gas was simulated by including reactions for species with ionization
potentials less than that of H (13.6 eV). For want of more specific
information, rates corresponding to the mean interstellar radiation field
in the solar neighbourhood (Draine, 1978) were adopted. Photoionization
processes are not usually supposed to affect the gas in HH outflows; it is
tempting to assume that HH objects are shielded from ionizing photons by
the dust in the molecular clouds in which they reside.  However, our
previous work (Giannini et al., 2004) has shown that the fits to the
observed {\h} emission can sometimes be improved by including
photoionization by the interstellar background UV radiation field.
Accordingly, we investigated systematically whether including
photoionization reactions improved the fit of the models to the
observations. It was
found that including photoionization reactions improved the fits to the
observations in all cases.

\subsection{Comparison with observations}\label{modres}

The present observations of emission lines of {\h} alone are usually
insufficient to determine uniquely the shock parameters, notably the shock
speed and age and the pre-shock gas density; measurements of the (weaker)
lines from levels of still higher excitation, particularly from higher
vibrational manifolds, would be required to do so. However, when other
observational constraints are taken into account, the range of possible
models is more closely circumscribed. First, previous echelle spectroscopy
by Davis et al. (1999) provides some constraints on the shock velocity for
the knots HH99B1 and HH99B3. Second, it seems reasonable to require that
the pre-shock parameters for all the knots which are physically associated
should be similar. Third, the models of the {\h} emission should be
compatible with the J-type shock models which reproduce the [{\fe}] and
[{\ci}] emission lines. When these additional factors are taken into
account, as described in Sections \ref{h2}, \ref{feemiss} and \ref{h2feemiss}
below, we
find that the models highlighted in Table~\ref{radmod} account best for
the {\h} emission observed towards HH99.

\subsubsection{{\h} emission}\label{h2}

The HH99B bow shock complex is physically connected, and it is reasonable
to expect similar pre-shock conditions in all the models of the associated
knots. Models with a pre-shock gas density $n_\mathrm{H} = 10^4$
cm$^{-3}$, which included photoionization reactions, were found to yield
the most consistent fits to the observations of all the HH objects in this
complex. The shock velocity was deduced to be about 30 km s$^{-1}$ for all
the objects, and the evolutionary ages derived from the models of B0, B2
and B3 were about 120 yr (see Table~\ref{radmod}). The observations of B1
from slits 1 and 2 were modelled separately from each other and yielded a
similar age (but a greater age than for the other objects).  HH99B1 forms
part of the head and one wing of the putative bow shock (see Figure
\ref{HH99:fig}); it is the only part of the bow shock whose emission was
measured along the wing (in slit 1), rather than across the head.  In
addition, the portion of HH99B1 captured in slit 2 lies further back from
the apex of the bow shock than any other object (see Figure
\ref{HH99:fig}).  The degree of excitation of {\h} is expected to
decrease along the wing, with increasing
distance from the apex, and so it is anticipated (and found) that HH99B1
exhibits a lower degree of excitation than the other associated knots.

\begin{figure}
\centering
\includegraphics[width=9cm]{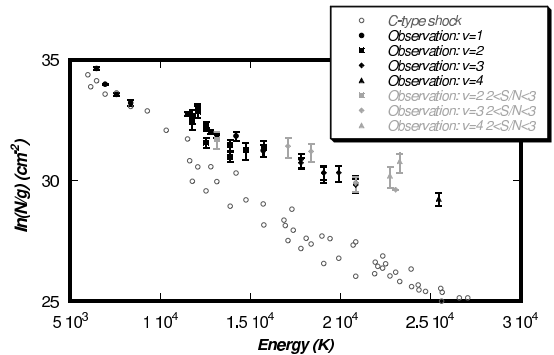}
\caption{The excitation diagram calculated in a steady-state C-type shock
(open circles) underestimates the observed intensities (filled symbols) of
high excitation lines.}
\label{Cshock}
\end{figure}

The importance of the J-type component of the models, to which we refer as
``J-type with magnetic precursor'', may be seen in Fig. (\ref{Cshock}).
There we
compare the excitation diagram which derives from the observations of
HH99B3 with that calculated in the limit of steady-state, when a pure
C-type shock wave obtains. It may be seen that the steady-state structure
underestimates the intensities of the H$_2$ lines of high excitation energy
and that the discrepancies increase with the excitation energy. A similar
conclusion was reached by Flower et al. (2003) when modelling the outflow
Cepheus A West.

A further constraint on $v_\mathrm{s}$ in HH99B1 and HH99B3 can be
derived from the echelle spectroscopy of Davis et al. (1999).  Using the
H$_2$ 2.122 $\mu$m line, they measured a FWHM of 15--20 km s$^{-1}$ in B1
and 20--40 km s$^{-1}$ in B3. These observations provide some indication
of the flow speed of the neutral gas in the region of the shock in which
the lines are excited; but it should be noted that the observed line
profiles were not deconvolved from the instrumental profile and that there
are (uncertain) line--of--sight projection effects. We note only that the
shock velocity of the model for HH99B3 is in good agreement with the
measured FWHM, whilst the narrower profile observed in B1 favours the
lower velocity model (i.e. $v_\mathrm{s}$ = 30 km s$^{-1}$) of the two
considered for this knot.

Davis et al. (1999) suggested that HH99A could be either the edge of the
HH99B bow shock or a clump of gas which has been overrun by the bow shock.
In the latter case, HH99A should have a greater evolutionary age than
HH99B.  Alternatively, HH99A could be the result of a separate outflow
event.  Referring to Table~\ref{radmod}, we see that all three possible
models of HH99A, labelled 1, 2 and 3 in the Table, are more evolved than
the corresponding models of HH99B; this is consistent with HH99A being a
clump of gas which has been overrun.

\begin{figure*}
\centering
\includegraphics{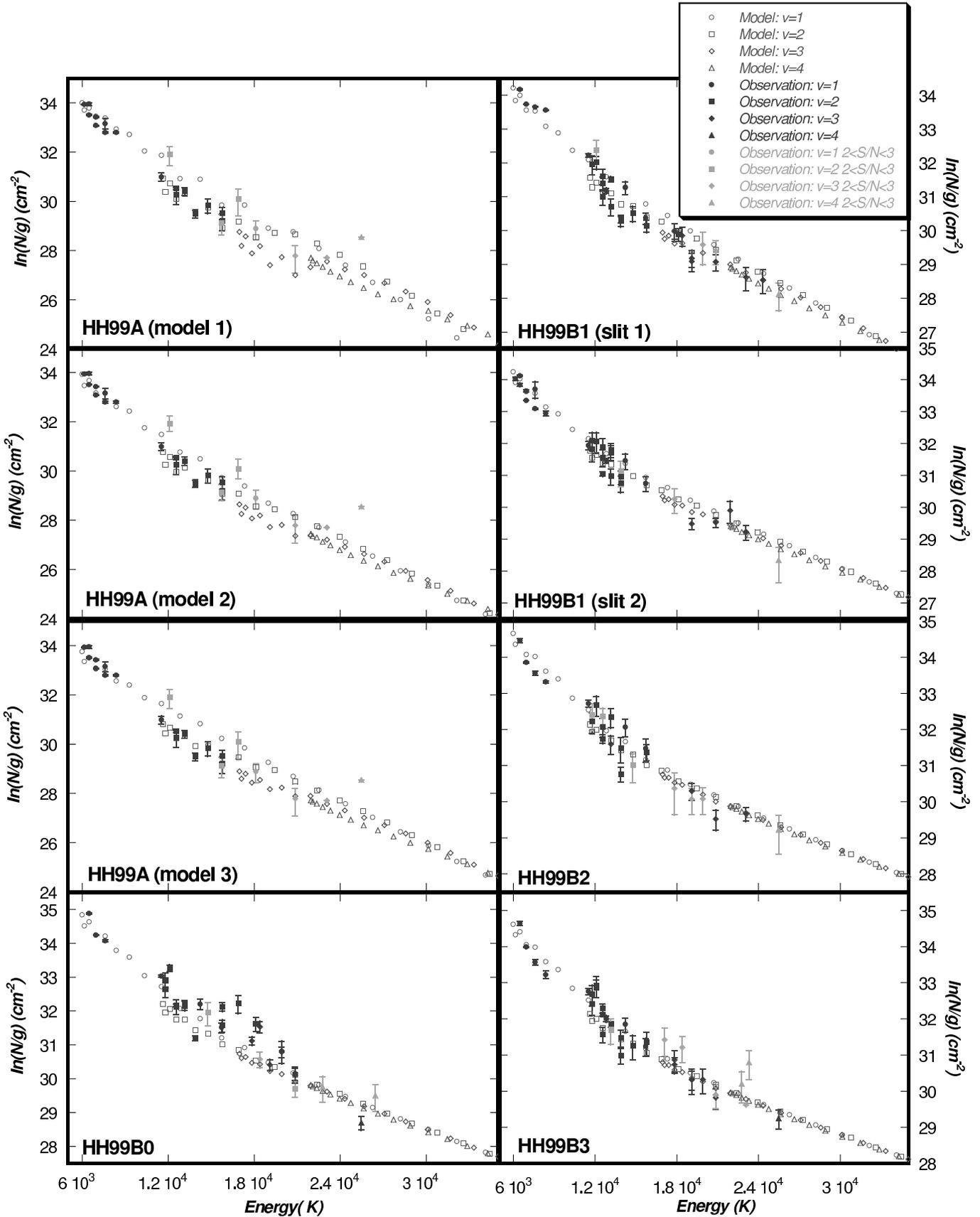}
\caption{Observed and computed excitation diagrams for HH99.  Transitions
from the $v = 1$, $v = 2$, $v = 3$ and $v = 4$ vibrational
manifolds are denoted by circles, squares, diamonds and triangles,
respectively. Open symbols represent the
models and filled symbols the observational results, with S/N$>$ 3
(black) and 2 $<$S/N$<$ 3 (grey).}
\label{H2ex}
\end{figure*}

The {\h} emission lines observed towards HH99A arise from levels with
excitation energies up to approximately $2.5\times 10^4$ K. However,
observations with S/N $> 3$ extend only up to $1.5\times 10^4$ K, as
compared with 2--2.5$\times 10^4$ K for the HH99B objects.  The lower S/N
of the lines of high excitation from HH99A compromises attempts to refine
the models further. The excitation diagrams derived from models (open
symbols) and observations (corresponding filled symbols) are shown in
Figure~\ref{H2ex}.  We plot the results for models 1--3 of HH99A and the
models adopted for the HH99B bow shock complex.

From the angular distance between HH99A and the HH99B complex ($25 \arcsec
$, see Figure~\ref{HH99:fig}) and the ages of our shock models, and
assuming a distance to HH99 of $d = 130$ pc (Marraco \& Rydgren, 1981), a
jet speed can be estimated. Taking 120 years as a typical age for the main
body of the HH99B complex, and the age of 350 yr from model 2 of HH99A, we
find a jet speed of 70 km s$^{-1}$.  This value is similar to the speeds
of 80--120 km s$^{-1}$ derived by Davis et al. (1999) from bow shock
models (which provided kinematic and spectroscopic data but did not
consider the chemistry and atomic processes included in the models
presented here) of the HH99 system.

\begin{table}
\caption{Parameters of possible shock models derived by fitting the {\h}
emission from the HH99A and the HH99B complex. Models considered to be the
best for each object are in bold face.}
\label{radmod}
\begin{tabular}{lccc}
\hline\\[-5pt]
Object & $v_\mathrm{s}$ & $n_{\mathrm{H}}^{a}$ & Age \\
\hline\\[-5pt]
& km s$^{-1}$ & cm$^{-3}$ & yr \\
\hline\\[-5pt]
HH99A(model 1)      & 50 & 10$^3$ & $>$ 3$\times$10$^3$ \\
HH99A(model 2)      & \bf{34} & \bf{5$\times$10$^3$} & \bf{350} \\
HH99A(model 3)      & 28 & 10$^4$ & 250 \\
\\[-5pt]
HH99B0         & \bf{34} & \bf{10$^4$} & \bf{130} \\
HH99B0         & 18 & 10$^5$ & 40  \\
\\[-5pt]
HH99B1(slit 1) & 55 & 10$^3$ & 400 \\
HH99B1(slit 1) & \bf{30} & \bf{10$^4$} & \bf{190} \\
\\[-5pt]
HH99B1(slit 2) & 53 & 10$^3$ & 200 \\
HH99B1(slit 2) & \bf{30} & \bf{10$^4$} & \bf{165} \\
\\[-5pt]
HH99B2         & 55 & 10$^3$ & 200 \\
HH99B2         & \bf{33} & \bf{10$^4$} & \bf{125} \\
\\[-5pt]
HH99B3         & \bf{33} & \bf{10$^4$} & \bf{115} \\
\hline\\[-5pt]
\end{tabular}\\
$^{a}$ $n_{\mathrm H}$ = $n$(H) + 2$n$(H$_{2}$)\\
\end{table}

\begin{table*}
\caption{Comparison between observed (reddening corrected, see Table for
$A_\mathrm{v}$ values) and computed [{\fe}] and [{\ci}] line intensities in
HH99}
\label{fluxtab}
\begin{tabular}{ccccc}
\hline\\[-5pt]
Object & Term & $\lambda$($\mu$m) & Observed (I $\pm$ $\Delta$I)  & Model \\
 & &  & \multicolumn{2}{c} { (10$^{-5}$ erg cm$^{-2}$ s$^{-1}$ sr$^{-1}$)} \\
\hline\\[-5pt]
HH99A$^1$ & [{\fe}]\,$a^4\!D_{7/2}-a^6\!D_{9/2}$ & 1.257 &  11.5$\pm$0.34 &
11.5 \\
$A_\mathrm{v}$=4.7 & [{\fe}]\,$a^4\!D_{7/2}-a^6\!D_{7/2}$ & 1.321 &
3.83$\pm$0.37 & 3.02 \\
 & [{\fe}]\,$a^4\!D_{7/2}-a^4\!F_{9/2}$ & 1.644 & 8.48$\pm$0.31 & 8.50 \\
 & [{\fe}]\,$a^4\!D_{5/2}-a^4\!F_{7/2}$ & 1.677 & 0.676$\pm$0.23 & 1.59 \\
 & [{\ci}]\,$^1\!D_{2}-^3\!P_{1}$ & 0.983 & 10.2$\pm$2.0 & 6.71 \\
 & [{\ci}]\,$^1\!D_{2}-^3\!P_{2}$ & 0.985 & 30.1$\pm$1.9 & 19.9 \\
\\[-5pt]
HH99B0 & [{\fe}]\,$a^4\!D_{7/2}-a^6\!D_{9/2}$ & 1.257 & 40.8$\pm$0.34 &
40.8 \\
$A_\mathrm{v}$=4.4 & [{\fe}]\,$a^4\!D_{1/2}-a^6\!D_{1/2}$ & 1.271 &
2.72$\pm$0.57 & 3.45 \\
 & [{\fe}]\,$a^4\!D_{3/2}-a^6\!D_{3/2}$ & 1.279 & 3.49$\pm$0.97 & 5.80 \\
 & [{\fe}]\,$a^4\!D_{5/2}-a^6\!D_{5/2}$ & 1.295 & 5.42$\pm$0.59 & 8.18 \\
 & [{\fe}]\,$a^4\!D_{3/2}-4^6\!D_{1/2}$ & 1.298 & 1.53$\pm$0.33 & 2.54 \\
 & [{\fe}]\,$a^4\!D_{7/2}-a^6\!D_{7/2}$ & 1.321 & 13.0$\pm$0.57 & 10.7 \\
 & [{\fe}]\,$a^4\!D_{5/2}-a^6\!D_{3/2}$ & 1.328 & 3.23$\pm$0.63 & 4.97 \\
 & [{\fe}]\,$a^4\!D_{5/2}-a^4\!F_{9/2}$ & 1.534 & 5.01$\pm$0.39 & 8.68 \\
 & [{\fe}]\,$a^4\!D_{3/2}-a^4\!F_{7/2}$ & 1.600 & 2.70$\pm$0.52 & 6.76 \\
 & [{\fe}]\,$a^4\!D_{7/2}-a^4\!F_{9/2}$ & 1.644 & 31.9$\pm$0.36 & 30.0 \\
 & [{\fe}]\,$a^4\!D_{1/2}-a^4\!F_{5/2}$ & 1.664 & 1.91$\pm$0.71$^a$ & 3.38 \\
 & [{\fe}]\,$a^4\!D_{5/2}-a^4\!F_{7/2}$ & 1.677 & 2.87$\pm$0.56 & 6.31 \\
 & [{\fe}]\,$a^4\!D_{3/2}-a^4\!F_{5/2}$ & 1.711 & 1.19$\pm$0.41$^a$ & 1.77 \\
 & [{\fe}]\,$a^4\!D_{1/2}-a^4\!F_{3/2}$ & 1.745 & 1.84$\pm$0.43 & 1.68 \\
 & [{\fe}]\,$a^4\!D_{3/2}-a^4\!F_{3/2}$ & 1.798 & 1.40$\pm$0.59$^a$ & 3.06 \\
 & [{\ci}]\,$^1\!D_{2}-^3\!P_{1}$ + $^1\!D_{2}-^3\!P_{2}$ & 0.983+0.985 &
37.2$\pm$3.2 & 40.1 \\
\\[-5pt]
HH99B1 (slit 1) & [{\fe}]\,$a^4\!D_{7/2}-a^6\!D_{9/2}$ & 1.257 &
9.36$\pm$0.35 & 7.64 \\
$A_\mathrm{v}$=6.7 & [{\fe}]\,$a^4\!D_{5/2}-a^4\!F_{9/2}$ & 1.534 &
2.01$\pm$0.36 & 1.66 \\
 & [{\fe}]\,$a^4\!D_{7/2}-a^4\!F_{9/2}$ & 1.644 & 7.06$\pm$0.70 & 5.62 \\
 & [{\ci}]\,$^1\!D_{2}-^3\!P_{1}$ + $^1\!D_{2}-^3\!P_{2}$ & 0.983+0.985 &
25.3$\pm$8.3 & 61.1 \\
\\[-5pt]
HH99B1 (slit 2)
 & [{\fe}]\,$a^4\!D_{7/2}-a^6\!D_{9/2}$ & 1.257 & 35.7$\pm$1.7 & 35.9 \\
$A_\mathrm{v}$=6.9 & [{\fe}]\,$a^4\!D_{7/2}-a^6\!D_{7/2}$ & 1.321 &
8.95$\pm$0.13 & 9.41 \\
 & [{\fe}]\,$a^4\!D_{5/2}-a^4\!F_{9/2}$ & 1.534 & 3.13$\pm$0.13$^a$ & 7.88 \\
 & [{\fe}]\,$a^4\!D_{7/2}-a^4\!F_{9/2}$ & 1.644 & 27.2$\pm$0.12 & 26.4 \\
 & [{\fe}]\,$a^4\!D_{5/2}-a^4\!F_{7/2}$ & 1.677 & 2.84$\pm$0.12$^a$ & 5.73 \\
 & [{\fe}]\,$a^4\!D_{3/2}-a^4\!F_{5/2}$ & 1.711 & 3.89$\pm$0.12 & 1.61 \\
 & [{\fe}]\,$a^4\!D_{1/2}-a^4\!F_{3/2}$ & 1.745 & 3.90$\pm$0.12 & 1.52 \\
 & [{\ci}]\,$^1\!D_{2}-^3\!P_{1}$ + $^1\!D_{2}-^3\!P_{2}$ & 0.983+0.985 &
200$^b$ & 38.0\\
\\[-5pt]
HH99B2 & [{\fe}]\,$a^4\!D_{7/2}-a^6\!D_{9/2}$ & 1.257 & 71.1$\pm$1.8 & 72.7 \\
$A_\mathrm{v}$=6.3 & [{\fe}]\,$a^4\!D_{5/2}-a^6\!D_{5/2}$ & 1.295 &
6.18$\pm$2.1 & 14.5 \\
 & [{\fe}]\,$a^4\!D_{7/2}-a^6\!D_{7/2}$ & 1.321 & 21.3$\pm$1.8 & 19.0 \\
 & [{\fe}]\,$a^4\!D_{5/2}-a^6\!D_{3/2}$ & 1.328 & 3.69$\pm$1.2 & 8.84 \\
 & [{\fe}]\,$a^4\!D_{5/2}-a^4\!F_{9/2}$ & 1.534 & 10.3$\pm$1.9 & 15.4 \\
 & [{\fe}]\,$a^4\!D_{3/2}-a^4\!F_{7/2}$ & 1.600 & 4.31$\pm$1.5$^a$ & 11.9 \\
 & [{\fe}]\,$a^4\!D_{7/2}-a^4\!F_{9/2}$ & 1.644 & 56.1$\pm$1.5 & 53.5 \\
 & [{\fe}]\,$a^4\!D_{1/2}-a^4\!F_{5/2}$ & 1.664 & 6.45$\pm$1.6 & 5.88 \\
 & [{\fe}]\,$a^4\!D_{5/2}-a^4\!F_{7/2}$ & 1.677 & 11.0$\pm$3.0 & 11.2 \\
 & [{\fe}]\,$a^4\!D_{3/2}-a^4\!F_{5/2}$ & 1.711 & 4.96$\pm$2.3$^a$ & 3.13 \\
 & [{\ci}]\,$^1\!D_{2}-^3\!P_{1}$ + $^1\!D_{2}-^3\!P_{2}$ & 0.983+0.985 &
94$^b$ & 22.5\\
\\[-5pt]
HH99B3 & [{\fe}]\,$a^4\!D_{3/2}-a^6\!D_{5/2}$ & 1.248 & 7.31$\pm$1.6 & 0.256 \\
$A_\mathrm{v}$=10.3 & [{\fe}]\,$a^4\!D_{7/2}-a^6\!D_{9/2}$ & 1.257 &
45.7$\pm$2.0 & 46.7 \\
 & [{\fe}]\,$a^4\!D_{1/2}-a^6\!D_{1/2}$ & 1.271 & 15.0$\pm$2.5 & 3.93 \\
 & [{\fe}]\,$a^4\!D_{3/2}-a^6\!D_{3/2}$ & 1.279 & 12.2$\pm$2.3 & 6.62 \\
 & [{\fe}]\,$a^4\!D_{7/2}-a^6\!D_{7/2}$ & 1.321 & 24.5$\pm$3.1 & 12.2 \\
 & [{\fe}]\,$a^4\!D_{7/2}-a^4\!F_{9/2}$ & 1.644 & 36.8$\pm$0.80 & 34.4 \\
 & [{\ci}]\,$^1\!D_{2}-^3\!P_{1}$ + $^1\!D_{2}-^3\!P_{2}$ & 0.983+0.985 &
269$^b$ & 36.4\\
\\[-5pt]
\hline\\[-5pt]
\end{tabular}\\
\flushleft
Notes: $^1$ Intensities from model (2) \\
$^a$ 2$<$S/N$<$3\\
$^b$ Upper limit
\flushright
\end{table*}

\subsubsection{[{\fe}] and [{\ci}] emission}\label{feemiss}

The models of HH99 described in Section \ref{h2} above underestimate the
observed intensities of the [{\fe}] and [{\ci}] lines by several orders of
magnitude. Giannini et al. (2004) suggested that the atomic and ionic
emission may arise in regions of higher ionization, distinct from those
responsible for the {\h} emission: at the apex of a bow shock or in a
reverse shock, within the jet, for example. Pure J-type shocks can yield
the necessary degrees of ionization, notably of H. The combination of a
pure J-type shock and either a pure C-type shock, or a J-type shock with
magnetic precursor, may be viewed as an elementary representation of a bow
shock.  Accordingly, we have attempted to model the atomic and ionic
emission from HH99 with pure J-type shocks.  For brevity, we shall refer
to the models in the previous Section as `{\h} models' from here on.

Under `normal' interstellar conditions, Fe is strongly depleted from the
gas phase, and most of the elemental abundance of Fe ($3.23\times
10^{-5}$: Anders \& Grevesse, 1989) is to be found in the grains, probably
in the form of silicates. In shock waves, dust grains may be shattered in
grain--grain collisions (Jones, Tielens, Hollenbach \& McKee, 1994; Jones,
Tielens \& Hollenbach, 1996; Flower \& Pineau des For\^{e}ts, 2003).
However, this process breaks large grains into smaller grains, rather than
releasing elements such as Fe into the gas phase.  In a J-type shock
wave, the neutral and charged fluids flow with the same speed. On the
other hand, the velocity difference between the neutral and charged fluids
in C-type shock waves gives rise to collisions between neutrals, such as
CO, and charged grains which are sufficiently energetic to result in the
erosion of the refractory grain cores (May et al., 2000). Thus, if a
C-type shock wave, or a J-type shock with a magnetic precursor, has passed
previously through a given region, the fraction of Fe in the gas phase
could be significant (providing that there has been insufficient time for
grains to reform).  In dynamically young objects, such as those under
study, which are subject to episodic events on timescales of the order of
500--1000 yr, or even less (e.g. Reipurth \& Bally, 2001), we consider
that it is reasonable to treat the gas phase abundance of Fe as a
variable. Once in the gas phase, Fe is ionized rapidly to Fe$^+$ through
charge transfer reactions with ions, and the [Fe II] spectrum can be
excited, principally by electron collisions, in sufficiently hot gas
(Giannini et al., 2004).

We assume that the pure J-type shock propagates into gas with a similar
density to that upstream of the corresponding {\h} model. Furthermore, we
include photoionization reactions, although these are less significant in
the context of the J-type component. J-type shocks which are sufficently
energetic to cause appreciable ionization of H are highly dissociative of
{\h} and do not produce significant emission in {\h} over the range of
excitation energies covered by the observations. J-type shocks with
$v_\mathrm{s} \approx 50$ km s$^{-1}$ have peak temperatures of order
10$^5$~K.  The intensities of the [{\fe}] lines are
insensitive to changes in temperature above 10$^4$~K, because the levels
from which emission arises are close together at 1.1--1.3$\times 10^4$~K
above ground. Consequently, the [{\fe}] lines are more sensitive to the
abundance of Fe in the gas phase than to the shock velocity, once this is
sufficient to give rise to partial ionization of hydrogen. In the models
of J-type shocks, $v_\mathrm{s} = 50$ km s$^{-1}$ was adopted, together
with the pre-shock densities used in the {\h} models, to be found in Table
\ref{radmod}. Only the gas phase abundance of Fe differentiated the
models: see Table~\ref{perFe}.

\begin{table}
\caption{The percentage of the elemental abundance of Fe in the gas phase,
as adopted in the models of pure J-type shocks.}
\label{perFe}
\begin{tabular}{lc}
\hline\\[-5pt]
Object & Gas phase Fe \\
\hline\\[-5pt]
HH99A (model 1) & 100 \\
HH99A (model 2) & 12  \\
HH99A (model 3) & 5   \\
HH99B0          & 25  \\
HH99B1 (slit 1) & 1   \\
HH99B1 (slit 2) & 23  \\
HH99B2          & 71  \\
HH99B3          & 31  \\
\hline\\[-5pt]
\end{tabular}
\end{table}

\begin{figure}
\centering
\includegraphics[width=9cm]{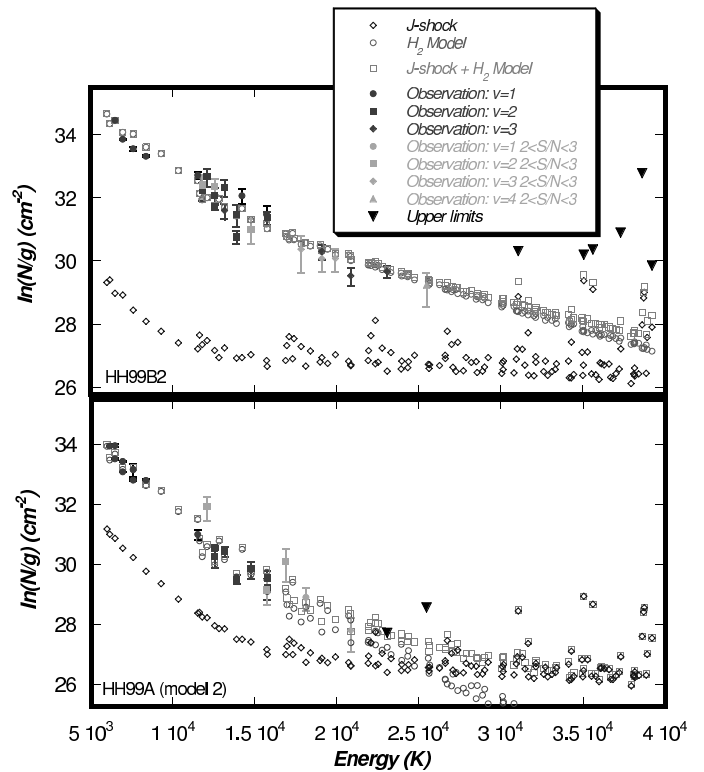}
\caption{Excitation diagrams from the {\h} models and pure J-type shock
components for HH99B2 (upper panel) and model 2 of HH99A (lower panel).  The
effect of the J-type shock is seen from the sum of the two components,
denoted ``J-shock + {\h} model''). The
contribution to the {\h} emission from the J-type shock is not significant
over the range of the observations of HH99B2 but becomes possibly
significant for
HH99A when the observational upper limits for transitions from levels with
excitation energies above 2.2$\times$10$^4$ K are taken into account.}
\label{all}
\end{figure}

A comparison of the predicted and observed intensities of both [{\fe}] and
[{\ci}]
lines is presented in
Table~\ref{fluxtab}, where only unblended lines of [{\fe}] are included.
The observations are corrected for reddening, using the power law fit of
Cardelli, Clayton \& Mathis (1989) to the data of Rieke \& Lebofsky
(1985), and adopting $R$ = 3.1. The predictions of the model are in
reasonably good agreement with the observations; the intensities of most
of the lines are reproduced to within a factor of 2 (or are consistent
with the upper limits for [{\ci}] observed with slit 2). In particular,
transitions from the a$^4$D$_{7/2}$ term of Fe$^+$ are generally
reproduced well, especially the strong 1.257 and 1.644 $\mu$m lines.  On
the other hand, the intensity of the 1.248 $\mu$m line (from the
a$^4$D$_{3/2}$ term) observed in HH99B3 is underestimated by more than an
order of magnitude.  It is possible that this line is partially blended
with the nearby 1.247 $\mu$m {\h} 2-0 Q(3) line, which is comparable or
stronger in intensity.  Although these lines show distinct peaks in the
spectrum, they are separated by only 9 $ \rm \AA$ and the de-blending
process may not be completely accurate for such a small separation.

The intensities of the [{\ci}] lines are reproduced without modification
of the pre-shock abundance of C in the gas phase.  Approximately 1/3 of
the elemental C abundance is already in the gas phase, in the form of CO,
which is dissociated at high temperature in reactions with H, producing C
and OH.  At the peak temperature of these J-type shocks, 99$\%$ of the
{\h} has been dissociated to H, and nearly all of the CO to C.

\subsubsection{{\h} emission from the [{\fe}] and [{\ci}]
 emitting region}\label{h2feemiss}

In the upper panel of Figure~\ref{all}, we plot the {\h} excitation diagrams
deriving from the J-type shock model and the {\h} model of HH99B2,
together with their sum.  It can be seen that the contribution to the {\h}
emission from the J-type shock is not significant over the observed range
of excitation energies; only above 3$\times 10^4$ K does the {\h} emission
from the J-type shock become important.  The shock wave is sufficiently
narrow that not all of the excited levels have time to thermalize, and
this results in the scatter about the median curve that can be seen in the
upper panel of Figure~\ref{all}.  Observational upper limits for HH99B2
are also plotted in the upper panel of Figure~\ref{all}; they fall well
above the model predictions. Similar results were obtained for the other
models of HH99B models and are not shown here.  The {\h} emission observed
towards HH99A exhibits lower excitation than is seen in the bow shock
complex. In the case of HH99A, the emission from the J-type shock becomes
significant above 2.2$\times 10^4$ K (see Figure~\ref{all}, lower panel).
This contribution provides a means of discriminating between the {\h}
models of HH99A.  When the contribution from the J-type shock is added to
that from models 1 and 3 ($n_\mathrm{H}$ = 10$^3$ and 10$^4$ cm$^{-3}$,
respectively), the upper limit to the column density of the $v = 3$ level
at 2.3$\times$ 10$^4$ K is exceeded.  In the case of model 2, this upper
limit is not violated.

\begin{figure}
\centering
\includegraphics[width=9cm]{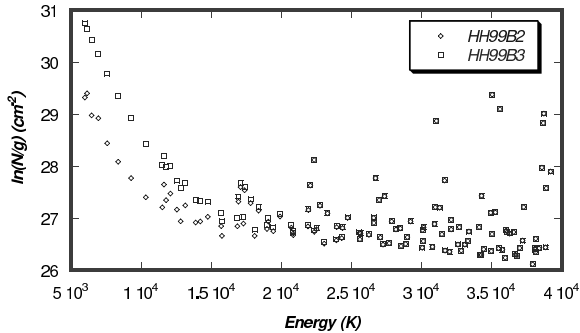}
\caption{The percentage of Fe in the gas phase
is the only parameter which changes between these models of HH99B2 (71$\%$)
and HH99B3 (31$\%$).
The effect of the different rates of cooling by Fe$^+$ in the post-shock
gas is reflected in the {\h} $v = 1$
level populations.}
\label{Fecool}
\end{figure}

The rapid and substantial dissociation of {\h} at the peak of these J-type
shock models reduces significantly the rate of cooling by {\h}, and
atomic/ionic cooling, for example of Fe$^+$ and O, following electron
collisional excitation, can become important.  The effect of cooling
by Fe$^+$ is seen in the postshock gas, in the region where the
temperature drops below 10$^4$~K.  At such temperatures, only the $v=1$
levels of {\h} are significantly populated; the
higher vibrational levels are populated immediately behind the
shock discontinuity.  The effect of the cooling due to Fe$^+$,
which becomes more
pronounced as the gas phase abundance of Fe$^+$ increases, is to reduce
the populations of the {\h} $v=1$ energy levels. By way of an
illustration, we compare, in Figure~\ref{Fecool}, the excitation diagrams
from the J-type shock models of B2 and B3. The only difference between the
models is the fraction of Fe in the gas phase (71$\%$ for HH99B2 and
31$\%$ for HH99B3). It may be seen that the higher gas phase abundance of
Fe in the model of HH99B2 results in lower populations of the $v = 1$
energy levels. It should be noted that this effect is not
observationally significant: the contribution of the pure J-type shock
to the emission from the {\h} $v=1$ levels  is insignificant compared with that
of the J-type shock with magnetic precursor.

\section{Conclusions}\label{Conc}

Spectra in the near IR (0.98--2.5 $\mu$m) of HH99 have been presented,
which are characterized by prominent emission in both [{\fe}] and
{\h} transitions. The strongest lines are observed at the head of the bow,
where emission in other atomic and ionic lines, of [S~II], [N~I] and
He~I, is also observed.  The observed intensities of the {\h}, [{\fe}]
and [{\ci}] emission lines have been used to constrain the parameters of
shock models.
Our conclusions are as follows:
\begin{itemize}

\item[-]{ with the possible exception of one model of HH99A, all the
models which fit the {\h} emission are of J-type shocks with magnetic
precursors; this implies that HH99 has not yet reached equilibrium, which
is consistent with outflows being young phenomena;}

\item[-]{ observations of {\h} lines from higher energy levels are required
to constrain further the {\h} models, of HH99A in particular;}

\item[-]{ the intensities of the [{\fe}] and [{\ci}] emission lines are
reproduced satisfactorily by a pure J-type shock with a velocity $v_s =
50$ km s$^{-1}$, which dissociates molecular hydrogen and partially
ionizes atomic hydrogen. In particular, the computed intensities of the
[Fe II] 1.257 $\mu$m and 1.644 $\mu$m lines, the strongest of the observed
transitions, are in good agreement with the observations;}

\item[-]{ in J-type shocks, Fe$^+$ becomes a significant coolant in the
post-shock gas (owing to dissociation of {\h} to H) and this has
consequences for the populations of the $v = 1$ levels of the residual
{\h}.}

\end{itemize}

Overall, our results are consistent with the conclusions that outflows are
episodic phenomena and that the emission from HH objects arises in bow
shocks, with the atomic and ionic emission lines being produced at the
apex and the {\h} emission lines in the wings.\\

\noindent \emph{Acknowledgments}\\
We thank an anonymous referee for a prompt and helpful report.

\end{document}